\documentclass[10pt,conference, letterpaper]{IEEEtran}
\IEEEoverridecommandlockouts
\usepackage{cite}
\usepackage{amsmath,amssymb,amsfonts}
\usepackage{algorithmic}
\usepackage{graphicx}
\usepackage{textcomp}
\usepackage{xcolor}
\def\BibTeX{{\rm B\kern-.05em{\sc i\kern-.025em b}\kern-.08em
    T\kern-.1667em\lower.7ex\hbox{E}\kern-.125emX}}

\usepackage{bbding}
\usepackage{pifont}
\usepackage{wasysym}
\usepackage{amssymb}
\usepackage[left=0.68in,top=0.78in,right=0.68in,bottom=0.78in]{geometry}
\newcommand{\etal}{\textit{et al.}}
\newcommand{\ie}{\textit{i.e.,~}}

\newcommand{\aka}{\textit{a.k.a. }}

\makeatletter
\let\old@ps@headings\ps@headings
\let\old@ps@IEEEtitlepagestyle\ps@IEEEtitlepagestyle
\def\confheader#1{
	\def\ps@IEEEtitlepagestyle{
		\old@ps@IEEEtitlepagestyle
		\def\@oddhead{\normalfont\scriptsize\centering#1}%
	}%
}

\makeatother

\confheader{%
	This paper has been accepted for publication
	in the IEEE Global Communications Conference (GLOBECOM'19), 09 December-13 December 2019, Waikoloa, HI, USA
}

\begin{document}

\title{A Lightweight and Privacy-Preserving Authentication Protocol for Mobile Edge Computing\\
}

\author{
	\IEEEauthorblockN{Kuljeet Kaur\IEEEauthorrefmark{1}, Member, IEEE, Sahil Garg\IEEEauthorrefmark{1}, Member, IEEE, Georges Kaddoum\IEEEauthorrefmark{1}, Member, IEEE,\\ Mohsen Guizani\IEEEauthorrefmark{2}, Fellow, IEEE, and Dushantha Nalin K. Jayakody\IEEEauthorrefmark{3}, Senior Member, IEEE.}
	\IEEEauthorblockA{\IEEEauthorrefmark{1}Electrical Engineering Department, \'Ecole de technologie sup\'erieure, Montr\'eal, QC H3C 1K3, Canada.\\
		\IEEEauthorrefmark{2}Computer Science and Engineering, Qatar University, Qatar.\\
		\IEEEauthorrefmark{3}School of Computer Science and Robotics, National Research Tomsk Polytechnic University, Russia.\\}
	(e-mail: kuljeet.kaur@ieee.org, sahil.garg@ieee.org, georges.kaddoum@etsmtl.ca, mguizani@ieee.org, and\\ nalin.jayakody@ieee.org)
}
\maketitle

\begin{abstract}	
	With the advent of the Internet-of-Things (IoT), vehicular networks and cyber-physical systems, the need for real-time data processing and analysis has emerged as an essential pre-requite for customers' satisfaction. In this direction, Mobile Edge Computing (MEC) provides seamless services with reduced latency, enhanced mobility, and improved location awareness. Since MEC has evolved from Cloud Computing, it inherited numerous security and privacy issues from the latter. Further, decentralized architectures and diversified deployment environments used in MEC platforms also aggravate the problem; causing great concerns for the research fraternity. Thus, in this paper, we propose an efficient and lightweight mutual authentication protocol for MEC environments; based on Elliptic Curve Cryptography (ECC), one-way hash functions and concatenation operations. The designed protocol also leverages the advantages of discrete logarithm problems, computational Diffie-Hellman, random numbers and time-stamps to resist various attacks namely-impersonation attacks, replay attacks, man-in-the-middle attacks, etc. The paper also presents a comparative assessment of the proposed scheme relative to the current state-of-the-art schemes. The obtained results demonstrate that the proposed scheme incurs relatively less communication and computational overheads, and is appropriate to be adopted in resource constraint MEC environments.


\end{abstract}


\section{Introduction}

Mobile Edge Computing (MEC) has emerged as a promising platform, providing seamless data storage, computational and processing facilities at the edge of the network. In MEC, heterogeneous arrays of devices (such as laptops, access points, computers, tablets, switches, etc.) are involved in providing intended services to the end-users. More formally, MEC extends the notion of cloud computing facility closer to the ground \cite{kaur2018edge}. However, in contrast to the distant cloud computing platform, MEC enables real-time data processing and analysis with reduced latency, enhanced mobility, greater user experience, and improved location awareness \cite{xiao2018security}. With these inherent advantages, the provisioning of emerging technologies like the Internet of Things (IoT),  vehicular networks, cyber-physical systems (CPSs) has become a reality \cite{8675175, garg2018edge}. \\
\indent However, the greatest challenge associated with the wide-scale adoption of MEC is induced due to the underlying security and privacy issues. MEC devices (\aka edge devices) are mostly located closer to the users with limited installed security measures and support. Further, the decentralized architectures and diversified deployment environments often used in MEC environments, adversely impact the provisioning of centralized security measures. Hence, MEC devices are often susceptible to various attack vectors such as man-in-the-middle attacks (MITM), impersonation attacks, user traceability attacks, replay attacks, etc. Further, the information relayed over the open channel is prone to sniffing, tampering and replay; leading to severe privacy leakage issues. In order to circumvent the above-mentioned challenges, the installation of mutual authentication protocols in MEC environments can be a viable solution. These protocols enable mutual verification of the communicating parties before the transmission of confidential and sensitive information. They also facilitate the key exchange mechanism for preserving confidentiality and privacy\cite{2019arXiv190401168K, 2019arXiv190401171G}. \\
\indent The following section summarizes the most related schemes proposed in the literature along the lines. Almajali \etal \cite{almajali2018framework} proposed a comprehensive architecture with the aim of instilling security in IoT environments specifically using MEC. Furthermore, the authors also evaluated the existing state-of-the-art authentication protocols under the proposed architecture. Likewise, the primary focus of Jia \etal's \cite{jia2019provably} work was on data security and user privacy, particularly in the context of MEC. To attain the same, the authors designed an identity-based anonymous authentication-driven key agreement protocol based on Elliptic Curve Cryptography (ECC). However, the authors employed expensive bilinear pairing operations in their protocol design. Xia \etal~ in \cite{xiao2018security} focused on the caching mechanism used in MEC; wherein the authors were found quoting that MEC is susceptible to numerous attack vectors ranging from denial of service to impersonation attacks. Consequently, the authors investigated three attack models in MEC environments in the context of mobile offloading and caching procedures. Additionally, the authors also propounded an authentication scheme for mobile edge devices.\\
\indent Tasi and Lo \cite{tsai2015privacy} presented a distributed scheme for mobile devices accessing cloud computing services. In the designed scheme, the authors employed computationally expensive bilinear pairing cryptosystem. The authors also claimed that their scheme was resilient against privacy comprising attacks and supported mutual authentication, replay protection, forgery protection, user anonymity, and user intractability. However, Jiang \etal  \cite{jiang2018security}  falsified Tasi and Lo's notion to maintain adequate user privacy, and proved that the scheme was vulnerable to user impersonation attacks and lacked mutual authentication support. The improved variants of Tasi and Lo's scheme along with its security loopholes were also discussed in \cite{irshad2016improved, amin2016more, he2018efficient}. For instance, Irshad \etal \cite{irshad2016improved} used bilinear pairing operations and map-to-point hash functions to meet the claimed security requirements of the Mobile Cloud Computing (MCC) platform. Unfortunately, Xiong \etal \cite{xiong2017enhanced} claimed that Irshad \etal's scheme was computationally expensive and was prone to two designs flaws, \ie inability to achieve three-factor security and lack of user revocation and registration facility.  Thus, to solve the above-mentioned issues, Xiong \etal \cite{xiong2017enhanced}  designed a provably secure authentication scheme, specifically for mobile computing setups. More recently, Li \etal \cite{li2019aep} proposed a preserving-preserving scheme based on ECC for mobile devices. The authors claim that their scheme supports mutual authentication, preserves users' privacy and provides perfect forward secrecy. 

It is evident from the above discussion that different variants of authentication protocols have been proposed in the context of MEC and MCC environments. Nevertheless, the existing protocols are prone to different attack vectors and lead to higher communicational and computational overheads. Thus, this paper presents a lightweight and privacy-preserving mutual authentication protocol, especially for MEC environments.


\subsection{Contributions}
The key contributions of the work are listed below:
\begin{itemize}
	\item We propose an ECC-based privacy-preserving mutual authentication protocol that is resistant to several sophisticated attacks in MEC environments. The designed protocol also helps in the key exchange between the participating entities, \ie mobile users and MEC servers.
	\item We present a detailed security analysis of the proposed protocol and compare its features with the existing state-of-the-art.
	\item Another objective of the paper is to protect mobile edge devices in an open and vulnerable network; while keeping the associated overheads to a minimum. 
\end{itemize}

\subsection{Organization}
The rest of the paper is organized as follows. Section~\ref{sec:SystemModel} sketches the schematic overview of the MEC environment. Section~\ref{sec:ProposedProtocol} illustrates the proposed protocol. The security and performance evaluations are presented in Sections~\ref{sec:SecurityAnalysis} and \ref{sec:PerformanceEvaluation}, respectively. Finally, the paper is concluded in Section~\ref{sec:Conclusion}.

\section{System Model} \label{sec:SystemModel}
\begin{figure}
	\centering
	\includegraphics[scale=.6]{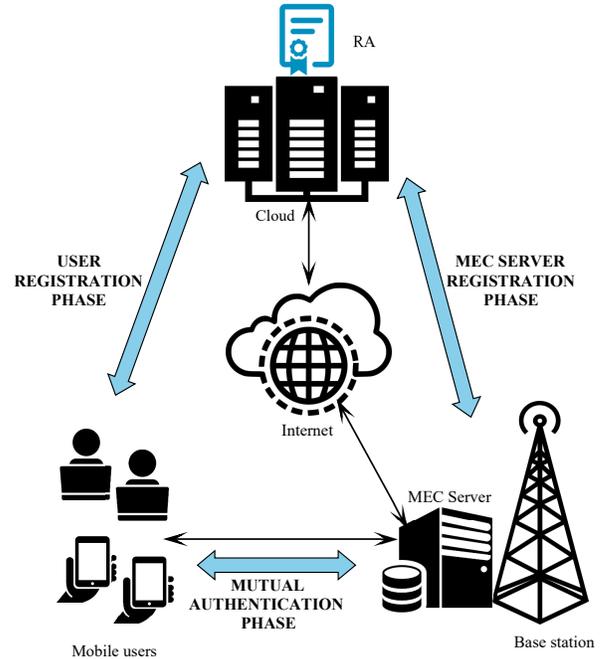}
	\caption{A typical setup of MEC.}
	\label{fig:systemmodel}
\end{figure}

A typical MEC environment consists of the following entities, \ie a trusted Registration Center (RC), mobile users, and MEC servers; as depicted in Fig.~\ref{fig:systemmodel}. Here, the MEC servers are provided with computational and storage functionalities. These servers are geographically dispersed and deployed in close proximity of the mobile users; often at mobile base stations. Their closeness to the users helps reduce the latency and enhance the user experience significantly. The mobile users can access MEC services via their vehicles, smartphones, tablets, etc.
On the contrary, the RC is assumed to be a trusted third-party that helps initialize the cryptographic parameters and provides a common platform for the registration of the users and MEC servers. The process of registration is essential for the proposed lightweight and privacy-preserving authentication protocol. In the designed protocol, it is assumed that the users and MEC servers get registered only once during their lifetime. Thus, the RC is deployed at the distant cloud platform and is not involved in the authentication process.

\subsection{Assumptions Considered}
The following list of assumptions have been considered in this work:
\begin{enumerate}
	\item It is assumed that the respective clocks of all the participating entities are well synchronized. This assumption enables generation of fresh timestamps for every session which helps the MEC resist to replay attacks.
	\item The threat model is considered to be Dolev-Yao (DY); under which an adversary ($\mathcal{A}$) has full access to the channel. This, in turn, provides the $\mathcal{A}$ with the ability to intercept, alter, replay and delete the transmitted messages. The DY threat model relates precisely with the practical open channels.
	\item The $\mathcal{A}$ can either be an insider or outsider.
	\item The channels i) between the mobile user and the RC and ii) between MEC server and the RC, are assumed to be secure.
\end{enumerate}

\section{Proposed Authentication Scheme} \label{sec:ProposedProtocol}
In this section, the proposed authentication protocol for MEC is presented. The designed protocol operates across the following phases: 1) Setup Phase, 2) User Registration Phase, 3) Server Registration Phase, and 4) Authentication Phase. 

\subsection{Phase I: Setup Phase} This phase forms the crux of the upcoming phases and initializes all the parameters essential for the security of the MEC environment. The related details are summarized below:

\vspace{0.7mm}
\noindent \textit{Step 1:} The RC begins the initialization process by selecting a cyclic additive group $G$,  an elliptic curve $E$ with a generator $P$ and order $q$. 

\vspace{0.7mm}
\noindent \textit{Step 2:} Next, the RC generates its public ($P_{RC}$) and private ($d_{RC}$) key-pairs as follows: $d_{RC} \in Z^{*}_q$ and $P_{RC}=d_{RC}.P$.

\vspace{0.7mm}
\noindent \textit{Step 3:} Next, a pair of collision-resistant one-way hash functions ($H_1()$ and $H_2()$) are chosen.

\vspace{0.7mm}
\noindent \textit{Step~4:}~Lastly, the following parameters $<G, E, P, q, H_1(),$ $H_2(), P_{RC}>$ are made public.

\subsection{Phase II: User Registration Phase}  During this phase, the $u^{th}$ mobile user registers itself with the RC and in return is allocated a dedicated public-private key pair ($P_u~\&~d_u$) and a pseudo-identity ($SID_u$). The detailed process is elaborated as under.

\vspace{0.7mm}
\noindent \textit{Step 1:} The $u^{th}$ mobile user selects an identity ($ID_u$) and transmits it to the RC over the secure channel.

\vspace{0.7mm}
\noindent \textit{Step 2:} Upon receiving the user's registration request, the RC generates a secret key $r_u$ and computes $R_u$. Following this, the RC performs the hash operation over the concatenated values of $ID_u$ and $R_u$, \ie $h_u=H_1(ID_u||R_u)$. Finally, the RC calculates $SID_u$ using the values computed above and its private key $d_{RC}$.

\vspace{0.7mm}
\noindent \textit{Step 3:} Then, the RC computes $P_u~\&~d_u$ for the $u^{th}$ user. The computed values, \ie $<SID_u, ~d_u,~r_u>$ are then transmitted to the user over the secure channel.

\vspace{0.7mm}
\noindent \textit{Step 4:} The received values and keys are stored by the user for the upcoming mutual authentication and key establishment phase. Meanwhile, the values  $<SID_u, ~P_u,~R_u>$ are made public to the MEC servers and stored in their respective repositories.

\begin{figure}[t]
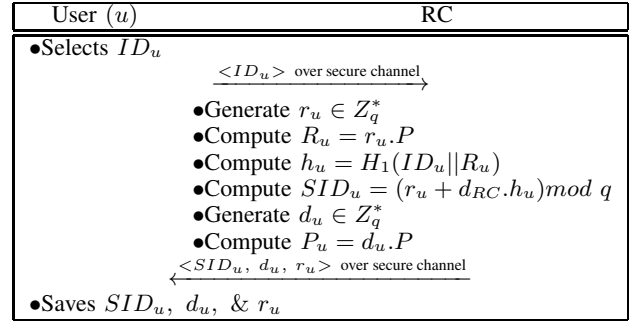

	\centering
	\small
	\begin{tabular}{|p{1.75cm} p{.5cm} p{4.4cm}|}
		\hline
		\multicolumn{1}{|c}{{User $(u)$}} & {}  & \multicolumn{1}{c|}{{RC}}   \\
		\hline
		\hline
		\multicolumn{3}{|l|}{$\bullet$Selects $ID_u$}\\
		\multicolumn{3}{|c|}{$\xrightarrow{<ID_u> ~\text{over secure channel}}$}\\
		&\multicolumn{2}{l|}{$\bullet$Generate $r_u \in Z^{*}_q$}\\
		&\multicolumn{2}{l|}{$\bullet$Compute $R_u=r_u.P$}\\
		&\multicolumn{2}{l|}{$\bullet$Compute $h_u=H_1(ID_u||R_u)$}\\
		&\multicolumn{2}{l|}{$\bullet$Compute $SID_u=(r_u+d_{RC}.h_u)mod~q$}\\
		&\multicolumn{2}{l|}{$\bullet$Generate $d_u \in Z^{*}_q$}\\
		&\multicolumn{2}{l|}{$\bullet$Compute $P_u=d_u.P$}\\
		\multicolumn{3}{|c|}{ $\xleftarrow{<SID_u, ~d_u,~r_u>~\text{over secure channel}}$}\\
		\multicolumn{3}{|l|}{$\bullet$Saves $SID_u, ~d_u,~\&~r_u$}\\
		
		\hline
	\end{tabular}
	\caption{Phase II: User Registration Phase.}
	\label{fig:Phase2}
\end{figure}

\subsection{Phase III: Server Registration Phase} 
This phase is identical to the previously discussed phase; wherein the only difference lies in the party trying to register with the RC. Here, the $ms^{th}$ MEC server tries to register itself and is associated with the following parameters: public-private key pairs ($P_{ms}~\&~d_{ms}$), pseudo-identity ($SID_{ms}$), and secret key ($r_{ms}$).
\begin{figure}[h]
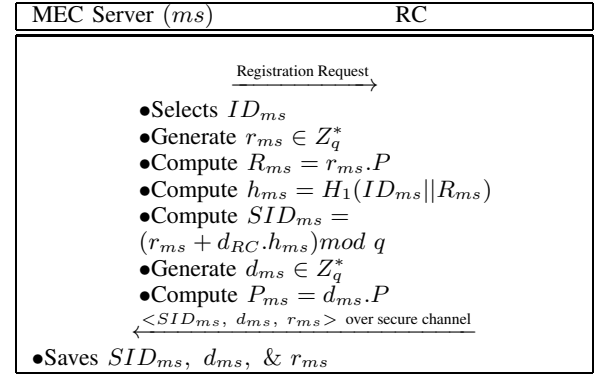

	\centering
	\small
	\begin{tabular}{|p{1.cm} p{1cm} p{4.4cm}|}
		\hline
		\multicolumn{2}{|c}{{MEC Server $(ms)$}}   & \multicolumn{1}{c|}{{RC}}   \\
		\hline
		\hline
		&&\\
		\multicolumn{3}{|c|}{$\xrightarrow{\text{Registration Request}}$}\\
		&\multicolumn{2}{l|}{$\bullet$Selects $ID_{ms}$}\\    
		&\multicolumn{2}{l|}{$\bullet$Generate $r_{ms} \in Z^{*}_q$}\\
		&\multicolumn{2}{l|}{$\bullet$Compute $R_{ms}=r_{ms}.P$}\\
		&\multicolumn{2}{l|}{$\bullet$Compute $h_{ms}=H_1(ID_{ms}||R_{ms})$}\\
		&\multicolumn{2}{l|}{$\bullet$Compute $SID_{ms}=$}\\
		&\multicolumn{2}{l|}{$(r_{ms}+d_{RC}.h_{ms})mod~q$}\\
		&\multicolumn{2}{l|}{$\bullet$Generate $d_{ms} \in Z^{*}_q$}\\
		&\multicolumn{2}{l|}{$\bullet$Compute $P_{ms}=d_{ms}.P$}\\
		\multicolumn{3}{|c|}{ $\xleftarrow{<SID_{ms}, ~d_{ms},~r_{ms}>~\text{over secure channel}}$}\\
		\multicolumn{3}{|l|}{$\bullet$Saves $SID_{ms}, ~d_{ms},~\&~r_{ms}$}\\
		\hline
	\end{tabular}
	\caption{Phase III: Server Registration Phase.}
	\label{fig:Phase3}
\end{figure}

\subsection{Phase IV: Authentication Phase} 
\begin{figure}[t]
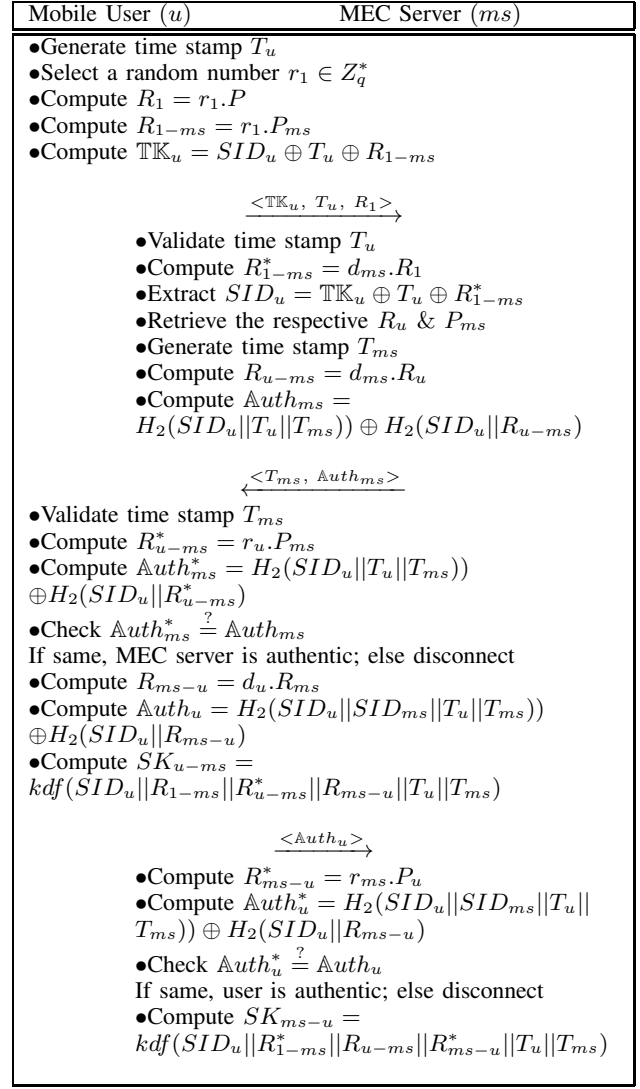

	\centering
	\small
	\begin{tabular}{|p{1cm} p{1cm} p{5cm}|}
		\hline
		\multicolumn{2}{|l}{{Mobile User $(u)$}} & \multicolumn{1}{c|}{{MEC Server $(ms)$}}   \\
		\hline
		\hline
		\multicolumn{3}{|l|}{$\bullet$Generate time stamp $T_u$ }\\
		\multicolumn{3}{|l|}{$\bullet$Select a random number $r_1\in Z^{*}_q$ }\\
		\multicolumn{3}{|l|}{$\bullet$Compute $R_1=r_1.P$ }\\
		\multicolumn{3}{|l|}{$\bullet$Compute $R_{1-ms}=r_1. P_{ms}$}\\
		\multicolumn{3}{|l|}{$\bullet$Compute $\mathbb{TK}_u = SID_u \oplus T_u \oplus R_{1-ms} $}\\
		
		&&\\
		\multicolumn{3}{|c|}{$\xrightarrow{<\mathbb{TK}_u,~T_u,~R_1>}$}\\
		
		&\multicolumn{2}{l|}{$\bullet$Validate time stamp $T_u$}\\
		&\multicolumn{2}{l|}{$\bullet$Compute  $R_{1-ms}^{*}=d_{ms}.R_1$}\\
		&\multicolumn{2}{l|}{$\bullet$Extract $SID_u = \mathbb{TK}_u \oplus T_u \oplus R_{1-ms}^{*}$}\\
		&\multicolumn{2}{l|}{$\bullet$Retrieve the respective $R_u ~\& ~P_{ms}$}\\
		&\multicolumn{2}{l|}{$\bullet$Generate time stamp $T_{ms}$}\\
		&\multicolumn{2}{l|}{$\bullet$Compute $R_{u-ms}= d_{ms}.R_u$}\\
		&\multicolumn{2}{l|}{$\bullet$Compute $\mathbb{A}uth_{ms} =$}\\
		&\multicolumn{2}{l|}{$H_2(SID_u||T_u|| T_{ms})) \oplus H_2(SID_u||R_{u-ms})$}\\
		&&\\
		\multicolumn{3}{|c|}{$\xleftarrow{<T_{ms}, ~\mathbb{A}uth_{ms}>}$}\\
		
		\multicolumn{3}{|l|}{$\bullet$Validate time stamp $T_{ms}$}\\
		\multicolumn{3}{|l|}{$\bullet$Compute $R_{u-ms}^{*}= r_u.P_{ms}$}\\
		\multicolumn{3}{|l|}{$\bullet$Compute $\mathbb{A}uth_{ms}^{*}= H_2(SID_u||T_u|| T_{ms}))$}\\
		\multicolumn{3}{|l|}{$\oplus H_2(SID_u || R_{u-ms}^{*} )$}\\
		\multicolumn{3}{|l|}{$\bullet$Check $\mathbb{A}uth_{ms}^{*} \stackrel{?}{=}  \mathbb{A}uth_{ms}$ } \\
		\multicolumn{3}{|l|}{If same, MEC server is authentic; else disconnect}\\
		\multicolumn{3}{|l|}{$\bullet$Compute $R_{ms-u}= d_u.R_{ms}$}\\
		\multicolumn{3}{|l|}{$\bullet$Compute $\mathbb{A}uth_{u}= H_2(SID_u||SID_{ms}||T_u|| T_{ms}))$}\\
		\multicolumn{3}{|l|}{$\oplus H_2(SID_u || R_{ms-u} )$}\\
		\multicolumn{3}{|l|}{$\bullet$Compute $SK_{u-ms} = $}\\
		\multicolumn{3}{|l|}{$kdf (SID_u ||  R_{1-ms} || R_{u-ms}^{*}||R_{ms-u} ||T_u || T_{ms} )$}\\
		&&\\
		\multicolumn{3}{|c|}{$\xrightarrow{<\mathbb{A}uth_{u}>}$}\\
		
		&\multicolumn{2}{l|}{$\bullet$Compute  $R_{ms-u}^{*}= r_{ms}.P_u$}\\
		&\multicolumn{2}{l|}{$\bullet$Compute $\mathbb{A}uth_{u}^{*}= H_2(SID_u||SID_{ms}||T_u|| $}\\
		&\multicolumn{2}{l|}{$ T_{ms})) \oplus H_2(SID_u || R_{ms-u} )$}\\
		&\multicolumn{2}{l|}{$\bullet$Check $\mathbb{A}uth_{u}^{*} \stackrel{?}{=}  \mathbb{A}uth_{u}$ } \\
		&\multicolumn{2}{l|}{If same, user is authentic; else disconnect}\\
		&\multicolumn{2}{l|}{$\bullet$Compute $SK_{ms-u} = $}\\
		&\multicolumn{2}{l|}{$kdf (SID_u ||  R_{1-ms}^{*} || R_{u-ms}||R_{ms-u}^{*} ||T_u || T_{ms} )$}\\
		&&\\
		\hline
		
	\end{tabular}
	\caption{Phase IV: Authentication Phase.}
	\label{fig:Phase4}
\end{figure}

During this phase, the mobile user and the MEC sever authenticate each other through a challenge-response mechanism and establish a shared session key for secure data transmission. The detailed processed is discussed as under (Fig.~\ref{fig:Phase4}).

\vspace{0.7mm}
\noindent \textit{Step 1:} The mutual authentication process is initiated by the mobile user trying to access the services of the MEC server. It initiates the process by recording its time-stamp ($T_u$) and generating a random number ($r_1$). Following this, the $u^{th}$ mobile user performs two ECC multiplication operations to generate $R_1$ and $R_{1-ms}$. Here, the latter is computed over $r_1$ and the $ms^{th}$ server's public key $P_{ms}$. 

\vspace{0.7mm}
\noindent \textit{Step 2:} Finally, the user computes a token $\mathbb{TK}_u$ for the MEC server to verify and extract its identity. The token is estimated using the XOR operation on $SID_u,~T_u,$ and  $R_{1-ms}$.

\vspace{0.7mm}
\noindent \textit{Step 3:} Next, the parameters $<\mathbb{TK}_u,~T_u,~R_1>$ are relayed to the MEC server.

\vspace{0.7mm}
\noindent \textit{Step 4:} Upon receiving these values, the MEC server validates the received time-stamp $T_u$. If the time-stamp lies within the permissible time window, then only the server advances the authentication phase. Otherwise, the connection is terminated.

\vspace{0.7mm}
\noindent \textit{Step 5:} Then, the server computes $R_{1-ms}^{*}=d_{ms}.R_1$ and extracts $SID_u$ using the XOR operations. Upon deriving a valid identity, the server checks its database for its corresponding cryptographic parameters $R_u ~\& ~P_{u}$.

\vspace{0.7mm}
\noindent \textit{Step 6:} The MEC serve then records its current time-stamp ($T_{ms}$) and computes $R_{u-ms}= d_{ms}.R_u$. Using these values, the $ms_{th}$ MEC server generates an intermediate authentication token ($\mathbb{A}uth_{ms}$) for the $u^{th}$ user to verify. Finally, the computed token and time-stamp are relayed to the user for verification.

\vspace{0.7mm}
\noindent \textit{Step 7:} Upon acquiring $<T_{ms}, ~\mathbb{A}uth_{ms}>$, the user first verifies the received $T_{ms}$. Next it computes $R_{u-ms}^{*}= r_u.P_{ms}$ using its secret key $r_u$; followed by the estimation of the authentication token $\mathbb{A}uth_{ms}^{*}= H_2(SID_u||T_u|| T_{ms})) \oplus H_2(SID_u || R_{u-ms}^{*} )$.

\vspace{0.7mm}
\noindent \textit{Step 8:} Finally, the tokens are checked for similarity. If the similarity is established then the user is assured that he is communicating with a valid MEC server; else the connection is terminated.

\vspace{0.7mm}
\noindent \textit{Step 9:} Next, the user computes $R_{ms-u}= d_u.R_{ms}$ and a token ($\mathbb{A}uth_{u}$) for the server to verify. Along with this, it also computes its shared session key ($SK_{u-ms}$) with the server; and it is used only if the mutual authentication phase is completed with positive outcomes.

\vspace{0.7mm}
\noindent \textit{Step 10:} Upon receiving $\mathbb{A}uth_{u}$, the $ms^{th}$ initiates the veri- -fication process by computing $\mathbb{A}uth_{u}^{*}$ and comparing it with the received value. If the two values match, then the process of mutual authentication between the two parties is established, and the server computes $SK_{u-ms}$.

\section{Security Analysis and Comparisons} \label{sec:SecurityAnalysis}

In this section, the security features provided by the proposed authentication scheme are discussed in details. In addition to this, the corresponding security evaluation with respect to the current state-of-the-art is also discussed herein.

\subsection{Security Analysis} 

The proposed authentication scheme is found to be resilient against a number of attack vectors and supports the following features:

\subsubsection{Mutual authentication} The proposed authentication mechanism supports mutual authentication between the mobile users and the MEC servers. The process of authentication is established by computing the intermediate authentication tokens, \ie $\mathbb{A}uth_{ms}$ and $\mathbb{A}uth_{u}$. These tokens can only be calculated by the legitimate users as they involve the knowledge of the secret keys $d_{ms}, r_u, d_u,$ and $r_{ms}$. Further, extracting the secret keys from their corresponding public keys in an intractable process in accordance with the elliptic curve discrete logarithm problem (ECDLP) \cite{kumar2016intelligent, jia2019provably}.

\subsubsection{Session key agreement} Phase IV of the proposed protocol helps establish shared secret key ($SK_{ms-u}, SK_{u-ms}$) between the participating entities. The computation of these keys requires the estimation of $R_{1-ms}, R_{1-ms}^{*}, R_{u-ms}, R_{u-ms}^{*}, R_{ms-u},$ and $R_{ms-u}^{*}$. Their computation in turn requires the knowledge of keys which are secret to the users and MEC servers. Further, the hardness of the computed $SK_{ms-u}$ and $SK_{u-ms}$ can be attributed to the elliptic curve diffie–hellman (ECDHP) problem \cite{jia2019provably}.
.

\subsubsection{User anonymity} In the designed protocol, the users' identity is completely masked using pseudo identity, \ie $SID_u$. Furthermore, neither the $ID_u$ nor $SID_u$ is ever transmitted to the server in clear text format. For instance, during the first authentication pass, $SID_u$ is masked using $T_u$ and $R_{1-ms}$. Here, the extraction of $R_{1-ms}$ is an infeasible process as it involves the information of  secret keys, \ie $r_1$ and $d_{ms}$.

\subsubsection{User intractability} The designed protocol guarantees un-traceability of mobile users as it involves the use of random numbers ($r_1$) and fresh time-stamps ($T_u~\&~T_{ms}$) in every session. Thus, the transmitted tokens ($\mathbb{TK}_u,~T_u,~R_1,~T_{ms}, ~\mathbb{A}uth_{ms},$ and $\mathbb{A}uth_{u}$) tend to acquire fresh values in each and every session. This limits $\mathcal{A}$'s capability to track a mobile user and monitor its activities.

\subsubsection{Single sign-in} In the considered set-up, the mobile user logins with the MEC server post successful mutual phase authentication and establishment of a shared session key. In the overall process, the mobile user needs to register itself with the RC only once. 

\subsubsection{Perfect forward secrecy} Let us assume that the $\mathcal{A}$ somehow gets hold of the private keys of the user and MEC server. Further, it is also assumed that $\mathcal{A}$ can now intercept all the ongoing information ($\mathbb{TK}_u,~T_u,~R_1,~T_{ms}, ~\mathbb{A}uth_{ms},$ and $\mathbb{A}uth_{u}$). Even under such a scenario, $\mathcal{A}$ would not be able to comprise the previously relayed messages without adequate knowledge of the random numbers ($r_1, r_u, r_{ms}$). Further, the computation of the session keys ($SK_{ms-u}$ and $SK_{u-ms}$) is a daunting task due to the nature of ECDHP problem.

\subsubsection{Resists impersonation attack} In order to successfully impersonate a user or MEC server, $\mathcal{A}$ needs the following information: $d_u, d_{ms}, r_1, r_{u}$ and $r_{ms}$. The said information is essential to generate the intermediate authentication tokens. Here, the first two in the list are the private keys and the last three are the random numbers and secret keys; and their extraction from their publicly available keys is an infeasible mechanism in accordance with ECDLP. 

\subsubsection{Resists  man-in-the-middle attack (MITM)} During this attack, $\mathcal{A}$ can intercept and alter the information transmitted over the channel between the two legitimate parties. Further, $\mathcal{A}$   can trick the legitimate users into believing that the modified message is being sent by an authorized source. In the proposed protocol, the intermediate tokens can only be changed with the help of private keys; which in turn can not be accessed by $\mathcal{A}$. Thus, the protocol is resilient against MITM attacks.

\subsubsection{Replay protection} The designed protocol also guarantees replay protection by incorporating the time-stamps ($T_u$ and $T_{ms}$) during each session. This means that the previously relayed messages cannot be used by the $\mathcal{A}$ to trick the legitimate users and servers into bypassing the mutual authentication process.


\subsection{Security Comparison}

An illustrative comparison of the proposed protocol against the existing schemes is detailed in Table~\ref{tb:ComparisionAuthentication}. The comparison is based on the following security features ($\mathfrak{SF}$): mutual authentication ($\mathfrak{SF}_1$), session key agreement ($\mathfrak{SF}_2$), user anonymity ($\mathfrak{SF}_3$), user intractability ($\mathfrak{SF}_4$),  single sign-in ($\mathfrak{SF}_5$), perfect forward secrecy ($\mathfrak{SF}_6$), resist impersonation attacks ($\mathfrak{SF}_7$), resist man-in-the-middle attacks ($\mathfrak{SF}_8$), replay protection ($\mathfrak{SF}_9$), and lightweight ($\mathfrak{SF}_{10}$). Three existing schemes namely Tsai \etal's \cite{tsai2015privacy}, Irshad \etal's \cite{irshad2016improved}, and Jia \etal's \cite{jia2019provably} schemes have been considered for the purpose of evaluation.

\begin{table}[ht]
	\centering
	\caption{Illustrative comparison of security features supported by the proposed and existing authentication protocols.}
	\label{tb:ComparisionAuthentication}
	\scriptsize
	\begin{tabular}{|p{1cm} | p{1.5cm}| p{1.65cm}|  p{1.4cm}|  p{1cm} |}
		\hline
		\textbf{Protocols} & Tsai \etal \cite{tsai2015privacy}  & Irshad \etal \cite{irshad2016improved} & Jia \etal \cite{jia2019provably}&  {Proposed} \\
		\hline
		\hline
		$\mathfrak{SF}_1$ & $\times$ &$\checkmark$&$\checkmark$&$\checkmark$ \\
		$\mathfrak{SF}_2$ & $\times$&$\checkmark$&$\checkmark$&$\checkmark$ \\
		$\mathfrak{SF}_3$ &$\times$&$\checkmark$&$\checkmark$&$\checkmark$ \\
		$\mathfrak{SF}_4$ &$\times$&$\checkmark$&$\checkmark$&$\checkmark$ \\
		$\mathfrak{SF}_5$ &$\checkmark$&$\checkmark$&$\checkmark$&$\checkmark$ \\
		$\mathfrak{SF}_6$&$\checkmark$&$\times$&$\checkmark$&$\checkmark$ \\
		$\mathfrak{SF}_7$ &$\times$&$\checkmark$&$\checkmark$&$\checkmark$ \\
		$\mathfrak{SF}_8$ &$\times$&$\checkmark$&$\checkmark$&$\checkmark$ \\
		$\mathfrak{SF}_9$ &$\checkmark$&$\checkmark$&$\checkmark$&$\checkmark$ \\
		$\mathfrak{SF}_{10}$ & $\times$&$\times$&$\times$&$\checkmark$ \\
		\hline
	\end{tabular}
	
\end{table}

In accordance with the details highlighted in Table~\ref{tb:ComparisionAuthentication} Tsai \etal's \cite{tsai2015privacy} scheme doesn't support the majority of the security features except for single sign-in, perfect forward secrecy and replay protection. Nevertheless, the protocols suggested by Irshad \etal~fail to support perfect forward secrecy and is not lightweight. Similarly, Jai \etal's scheme is also not lightweight but is secure against the major attack vectors. Thus, it is evident from the above discussion that the existing schemes/protocols cannot provide an adequate level of security and are not apt for MEC environments due to the use of heavy computations. Mobile devices involved in MEC often are battery constraint and have limited computational and communication capacity. Thus, heavy computations may take a toll on their performances. Meanwhile, the proposed authentication protocol supports all the security features and is lightweight too.

\section{Performance Evaluation} \label{sec:PerformanceEvaluation}
In this section, a comparative performance assessment of the proposed scheme with the existing schemes \cite{tsai2015privacy, irshad2016improved, jia2019provably} is demonstrated. These schemes are essentially identity-based and privacy-driven. The related simulation details are discussed next:

\subsection{Simulation Details}
The simulation setup is adopted from Jia \etal's \cite{jia2019provably} work; wherein the authors have adopted different platforms to emulate the computational capabilities of the MEC servers and mobile devices. Alibaba's cloud was used to simulate the servers with the following configurations: Intel(R)
Xeon(R) CPU E5-2630 0 @ 2.30 GHz, 1 GB RAM and Ubuntu 14.04. On the other hand, Google Nexus One smartphone provides the mobile devices platform with 2 GHz ARM CPU armeabi-v7a, 300 MiB RAM, and Android 4.4. For detailed information, readers are encouraged to refer to \cite{jia2019provably}. 

\subsection{Evaluation Parameters} The evaluation has been carried out on the basis of two widely used metrics, \ie computational overhead, and communication overhead.

\subsubsection{Computational Overhead Analysis}
For the computational analysis, different cryptographic functions were executed on the above-mentioned devices and platforms, \ie MEC servers and mobile devices. Table~\ref{tbl:ComputatonalProcessingTime} describes the overhead associated with the execution of different functions (in ms). Here, the variables $\mathbb{T}_{bp}, \mathbb{T}_{m}, \mathbb{T}_{a}, \mathbb{T}_{h},$ and $\mathbb{T}_{e}$ denote the computational time required for executing a single bilinear pairing, scalar multiplication, point addition, hashing, and modular exponentiation operation, respectively. 
\begin{table}[h]
	\centering
	\caption{Execution overhead of different cryptographic functions.}
	\label{tbl:ComputatonalProcessingTime}
	\begin{tabular}{|p{.9cm}| p{.95cm}| p{.95cm}| p{.95cm}| p{.95cm}|  p{.95cm}|}
		\hline
		
		Entity & $\mathbb{T}_{bp}$ & $\mathbb{T}_{m}$ & $\mathbb{T}_{a}$& $\mathbb{T}_{h}$ & $\mathbb{T}_{e}$  \\
		\hline
		\hline
		Server  & 5.275 & 1.97 & 0.012 & 0.009 & 0.339\\
		Client & 48.66 & 19.919 & 0.118 & 0.089 & 3.328 \\
		\hline
	\end{tabular}
\end{table}

Based on these values, the computational cost of the proposed protocol was analyzed against the existing protocols. The comparisons were carried out on both the mobile users' and MEC servers' sides. The obtained results are depicted in Table~\ref{tbl:ComputationOverhead}. It is evident from the results that the proposed scheme depicts the best performance for both the MEC servers and the mobile users, and it is the most lightweight protocol amongst all. 
\begin{table}[h]
	\centering
	
	\caption{Computational overhead analysis.}
	\label{tbl:ComputationOverhead}

		\begin{tabular}{|p{1.9cm}|p{2.95cm}|p{2.6cm}|}
			\hline
			
			\textbf{Protocols} & \textbf{At User} & \textbf{At Server }\\
			\hline
			\hline
	
			Tsai \etal \cite{tsai2015privacy}  &$5\mathbb{T}_{bp} + 2\mathbb{T}_{a}+ \mathbb{T}_{e} + \mathbb{T}_{inv} + 5 \mathbb{T}_{h} \approx 93.604$&$2\mathbb{T}_{bp} + 2\mathbb{T}_{m} + 2\mathbb{T}_{a}+ 2\mathbb{T}_{e}+5\mathbb{T}_{h} \approx 15.228$\\
			Irshad \etal \cite{irshad2016improved} &$\mathbb{T}_{bp} + 5\mathbb{T}_{m} + 2\mathbb{T}_{a}+ 2\mathbb{T}_{e} + \mathbb{T}_{inv} + 6\mathbb{T}_{h} \approx 155.681$& $2\mathbb{T}_{bp} + 4\mathbb{T}_{m} + 3\mathbb{T}_{a}+ 2\mathbb{T}_{e}+3\mathbb{T}_{h} \approx 19.171$\\
			Jia \etal \cite{jia2019provably} & $4\mathbb{T}_{m}+ 3\mathbb{T}_{a}+\mathbb{T}_{e}+5\mathbb{T}_{h} \approx 83.807$ & $\mathbb{T}_{bp}+5\mathbb{T}_{m} + 3\mathbb{T}_{a}+ 5\mathbb{T}_{h} \approx 15.206$\\
			Proposed & $4\mathbb{T}_{m} + 4\mathbb{T}_{h} \approx 80.032$ & $3\mathbb{T}_{m} + 4\mathbb{T}_{h} \approx 5.946$ \\
			\hline
		\end{tabular}
\end{table}

\subsubsection{Communication Overhead Analysis}
To analyze the communication overhead associated with the proposed and existing protocols, the number of messages transmitted between the participating parties has been considered as a potential metric. These transmitted messages comprise of different tokens such as cyclic additive group ($G$), field ($Z_q$), identity ($ID$), hash ($H$), and time-stamp ($T$). Their lengths were set as follows: $|G|= 1024 b, |Z_q|=160 b, |ID|=256 b, |H|=256 b,$ and $|T|=32 b$. In accordance with these values, the communication cost of the protocols was computed and summarized in Table~\ref{tbl:CommunicationOverhead}. The results depict that the proposed scheme leads to the best performance, \ie 2624 b followed by Tsai \etal's scheme (4608 b), Jia \etal's scheme (4736 b), and Irshad \etal's scheme (5632 b), respectively.

\begin{table}[h]
	\centering
	\caption{Comparison of communication overheads.}
	\label{tbl:CommunicationOverhead}
	\begin{tabular}{|p{2cm}| p{3.7cm}| p{1.2cm}|}
		\hline
		
		Scheme & Cost & No. of bits\\
		\hline
		\hline
		Tsai \etal \cite{tsai2015privacy} & $3|G|+|G_{T}|+|H|+|ID| $& 4608 \\
		Irshad \etal \cite{irshad2016improved}& $4|G|+|G_{T}|+|H|+|ID| $ &  5632\\
		Jia \etal \cite{jia2019provably} & $4|G|+2|{T}|+2|Z_{q}|+|ID| $ &  4736 \\
		Proposed &$2|G| + 2|T| +2|H|$ & 2624\\
		\hline
	\end{tabular}
\end{table}

\section{Conclusion} \label{sec:Conclusion}
In this paper, a lightweight and privacy-preserving mutual authentication protocol for MEC environments has been proposed. The proposed protocol relies on ECC, one-way hash functions and concatenation operations for its secure and lightweight functionality. More specifically, the designed protocol acquires its strength from the hardness of ECDLP and ECDHP problems to compute the intermediate authentication tokens. Additionally, the protocol also leverages the advantages of freshly generated random numbers and time-stamps to resist replays of previously communicated messages. Overall, the protocol withstands different security attacks (namely user impersonation, server impersonation, MITM, user traceability, etc.) and provides mutual authentication, session key agreement, and single sign-in facility. The performance assessment of the proposed protocols with the existing state-of-the-art protocol reveals that the proposed protocol achieves the highest level of security with comparatively reduced computational and communication overheads.

\section*{Acknowledgment}
This work was supported by the Tier 2 Canada Research Chair.

\bibliographystyle{IEEEtran}
\bibliography{references.bib}

\begin{thebibliography}{10}
\providecommand{\url}[1]{#1}
\csname url@samestyle\endcsname
\providecommand{\newblock}{\relax}
\providecommand{\bibinfo}[2]{#2}
\providecommand{\BIBentrySTDinterwordspacing}{\spaceskip=0pt\relax}
\providecommand{\BIBentryALTinterwordstretchfactor}{4}
\providecommand{\BIBentryALTinterwordspacing}{\spaceskip=\fontdimen2\font plus
\BIBentryALTinterwordstretchfactor\fontdimen3\font minus
  \fontdimen4\font\relax}
\providecommand{\BIBforeignlanguage}[2]{{%
\expandafter\ifx\csname l@#1\endcsname\relax
\typeout{** WARNING: IEEEtran.bst: No hyphenation pattern has been}%
\typeout{** loaded for the language `#1'. Using the pattern for}%
\typeout{** the default language instead.}%
\else
\language=\csname l@#1\endcsname
\fi
#2}}
\providecommand{\BIBdecl}{\relax}
\BIBdecl

\bibitem{kaur2018edge}
K.~Kaur, S.~Garg, G.~S. Aujla, N.~Kumar, J.~J. Rodrigues, and M.~Guizani,
  ``{Edge computing in the industrial internet of things environment:
  Software-defined-networks-based edge-cloud interplay},'' \emph{IEEE
  communications magazine}, vol.~56, no.~2, pp. 44--51, 2018.

\bibitem{xiao2018security}
L.~Xiao, X.~Wan, C.~Dai, X.~Du, X.~Chen, and M.~Guizani, ``{Security in mobile
  edge caching with reinforcement learning},'' \emph{IEEE Wireless
  Communications}, vol.~25, no.~3, pp. 116--122, 2018.

\bibitem{8675175}
S.~{Garg}, A.~{Singh}, K.~{Kaur}, G.~S. {Aujla}, S.~{Batra}, N.~{Kumar}, and
  M.~S. {Obaidat}, ``{Edge Computing-Based Security Framework for Big Data
  Analytics in VANETs},'' \emph{IEEE Network}, vol.~33, no.~2, pp. 72--81,
  2019.

\bibitem{garg2018edge}
S.~Garg, A.~Singh, K.~Kaur, S.~Batra, N.~Kumar, and M.~S. Obaidat,
  ``{Edge-based content delivery for providing QoE in wireless networks using
  quotient filter},'' in \emph{IEEE International Conference on Communications
  (ICC), Kansas City, USA}, May 2018.

\bibitem{2019arXiv190401168K}
K.~{Kaur}, S.~{Garg}, G.~{Kaddoum}, F.~{Gagnon}, and S.~H. {Ahmed},
  ``{Blockchain-Based Lightweight Authentication Mechanism for Vehicular Fog
  Infrastructure},'' in \emph{IEEE International Conference on Communications
  Workshops (ICC Workshops), Shanghai, China}, May 2019.

\bibitem{2019arXiv190401171G}
S.~{Garg}, K.~{Kaur}, G.~{Kaddoum}, F.~{Gagnon}, and J.~J. P.~C. {Rodrigues},
  ``{An Efficient Blockchain-Based Hierarchical Authentication Mechanism for
  Energy Trading in V2G Environment},'' in \emph{IEEE International Conference
  on Communications Workshops (ICC Workshops), Shanghai, China}, May 2019.

\bibitem{almajali2018framework}
S.~Almajali, H.~B. Salameh, M.~Ayyash, and H.~Elgala, ``{A framework for
  efficient and secured mobility of IoT devices in mobile edge computing},'' in
  \emph{Third International Conference on Fog and Mobile Edge Computing (FMEC),
  Barcelona, Spain}.\hskip 1em plus 0.5em minus 0.4em\relax IEEE, April 2018.

\bibitem{jia2019provably}
X.~Jia, D.~He, N.~Kumar, and K.-K.~R. Choo, ``{A Provably Secure and Efficient
  Identity-Based Anonymous Authentication Scheme for Mobile Edge Computing},''
  \emph{IEEE Systems Journal}, 2019, {}DOI: 10.1109/JSYST.2019.2896064.

\bibitem{tsai2015privacy}
J.-L. Tsai and N.-W. Lo, ``{A privacy-aware authentication scheme for
  distributed mobile cloud computing services},'' \emph{IEEE Systems Journal},
  vol.~9, no.~3, pp. 805--815, 2015.

\bibitem{jiang2018security}
Q.~Jiang, J.~Ma, and F.~Wei, ``{On the security of a privacy-aware
  authentication scheme for distributed mobile cloud computing services},''
  \emph{IEEE Systems Journal}, vol.~12, no.~2, pp. 2039--2042, 2018.

\bibitem{irshad2016improved}
A.~Irshad, M.~Sher, H.~F. Ahmad, B.~A. Alzahrani, S.~A. Chaudhry, and R.~Kumar,
  ``{An improved Multi-server Authentication Scheme for Distributed Mobile
  Cloud Computing Services},'' \emph{{KSII Transactions on Internet and
  Information Systems}}, vol.~10, no.~12, pp. 6092--6115, 2016.

\bibitem{amin2016more}
R.~Amin, S.~H. Islam, G.~Biswas, D.~Giri, M.~K. Khan, and N.~Kumar, ``{A more
  secure and privacy-aware anonymous user authentication scheme for distributed
  mobile cloud computing environments},'' \emph{Security and Communication
  Networks}, vol.~9, no.~17, pp. 4650--4666, 2016.

\bibitem{he2018efficient}
D.~He, N.~Kumar, M.~K. Khan, L.~Wang, and J.~Shen, ``{Efficient privacy-aware
  authentication scheme for mobile cloud computing services},'' \emph{IEEE
  Systems Journal}, vol.~12, no.~2, pp. 1621--1631, 2018.

\bibitem{xiong2017enhanced}
L.~Xiong, D.~Peng, T.~Peng, and H.~Liang, ``{An Enhanced Privacy-Aware
  Authentication Scheme for Distributed Mobile Cloud Computing Services},''
  \emph{KSII Transactions on Internet and Information Systems}, vol.~11,
  no.~12, pp. 6169--6187, 2017.

\bibitem{li2019aep}
J.~Li, W.~Zhang, V.~Dabra, K.-K.~R. Choo, S.~Kumari, and D.~Hogrefe,
  ``{AEP-PPA: An anonymous, efficient and provably-secure privacy-preserving
  authentication protocol for mobile services in smart cities},'' \emph{Journal
  of Network and Computer Applications}, vol. 134, pp. 52--61, 2019.

\bibitem{kumar2016intelligent}
N.~Kumar, K.~Kaur, S.~C. Misra, and R.~Iqbal, ``{An intelligent RFID-enabled
  authentication scheme for healthcare applications in vehicular mobile
  cloud},'' \emph{Peer-to-Peer Networking and Applications}, vol.~9, no.~5, pp.
  824--840, 2016.

\end{thebibliography}

\end{document}